\documentclass[aps,pra,reprint,amsmath,amssymb,superscriptaddress]{revtex4-1}
\usepackage{graphicx}
\usepackage{mathrsfs}
\renewcommand{\phi}{\varphi}
\usepackage{xfrac}
\usepackage[colorlinks]{hyperref}
\usepackage{braket}
\usepackage{mathrsfs}
\usepackage{times}
\usepackage[utf8]{inputenc}
\mathchardef\Re="023C
\mathchardef\Im="023D
\usepackage{subcaption}
\usepackage{slashed}

\usepackage{color}
\usepackage{tikz}
\usepackage{comment}

\makeindex
\begin{document}

\title{Fractionalized holes in one-dimensional $\mathbb{Z}_2$ gauge theory coupled to fermion matter ---\\ deconfined dynamics and emergent integrability}

\author{Aritra Das}
\affiliation{Indian Institute of Science, Bengaluru 560012, India}
\affiliation{Department of Physics, University of Maryland, College Park, Maryland 20742, USA}
\author{Umberto Borla}
\affiliation{Physik-Department, Technische Universität München, 85748 Garching, Germany}
\affiliation{Munich Center for Quantum Science and Technology (MCQST), 80799 München, Germany}
\author{Sergej Moroz}
\affiliation{Physik-Department, Technische Universität München, 85748 Garching, Germany}
\affiliation{Munich Center for Quantum Science and Technology (MCQST), 80799 München, Germany}
\affiliation{Department of Engineering and Physics, Karlstad University, Karlstad, Sweden}
\affiliation{Nordita, KTH Royal Institute of Technology and Stockholm University, Stockholm, Sweden}

\begin{abstract}
We investigate the interplay of quantum one-dimensional discrete $\mathbb{Z}_2$ gauge fields and fermion matter near full filling in terms of deconfined fractionalized hole excitations that constitute mobile domain walls between vacua that break spontaneously translation symmetry.
In the limit of strong string tension, we uncover emergent integrable correlated hopping dynamics of holes which is complementary to the constrained XXZ description in terms of bosonic dimers. We analyze numerically quantum dynamics of spreading of an isolated hole together with the associated time evolution of entanglement and provide analytical understanding of its salient features. We also study the model enriched with a short-range interaction and clarify the nature of the resulting ground state at low filling of holes and identify deconfined hole excitations near the hole filling $\nu^h=1/3$.
\end{abstract}

\maketitle

\section{Introduction} \label{Intro}
Identifying the origin of a linear attractive potential between charges mediated by gauge fields, known as confinement, and studying its consequences has been a long-standing challenge in nuclear, high-energy and condensed-matter physics \cite{greensite2011introduction, Fradkin2013, wenbook, mussardo2011integrability}. Recently, consequences of confinement on real-time quantum dynamics in spin chains and equivalent lattice gauge theories have been studied extensively \cite{kormos2017real, PhysRevB.99.180302, PhysRevB.102.014308, PhysRevB.102.041118, vovrosh2020confinement, PhysRevB.99.195108, PhysRevLett.122.130603, PhysRevLett.124.180602, magnifico2020real, PhysRevLett.122.150601, surace2021scattering, karpov2020spatiotemporal, lagnese2021false, milsted2021collisions, rigobello2021entanglement, Javier_Valencia_Tortora_2020, pomponio2021bloch, bastianello2021fragmentation, scopa2021entanglement}. Generically, confinement is known to hinder quantum thermalization and slow quantum dynamics by strongly suppressing the spreading of quantum correlations and entanglement growth. In the limit of strict confinement the Hilbert space can exhibit fragmentation \cite{PhysRevLett.124.207602, borla2020quantum, bastianello2021fragmentation} with emergent low-energy fracton excitations \cite{PhysRevResearch.2.013094, borla2020quantum}.

Here we consider one-component $U(1)$-symmetric quantum fermion matter hopping on a one-dimensional lattice coupled to dynamical $\mathbb{Z}_2$ gauge fields \cite{PhysRevB.84.235148, Barbiero2018, Schweizer2019, Frank_2020, borla2020confined, PhysRevB.101.024306, PhysRevLett.127.167203, PhysRevLett.125.030503, halimeh2021enhancing} that mediate attractive confining interaction between lattice fermions \footnote{Notwithstanding, at partial fermion filling the model is known to form a Luttinger liquid that exhibits gapless deconfined low-energy excitations. \cite{PhysRevB.84.235148, borla2020confined, PhysRevLett.127.167203}.}. In this paper we draw attention to and investigate the peculiar physics of the fractionalized holes which are domain walls between states fully filled with fermions. Due to the presence of the Ising gauge field, these vacua break translation symmetry spontaneously. Consequently, the holes are deconfined and interact via a long-range zig-zag potential in contrast to the confining linear interaction between the original fermions.  We study in detail the dynamics of one and two holes and provide several  independent manifestations of their deconfined nature. In the limit of strong string tension, we find that holes become heavy and undergo a peculiar correlated hopping of the type studied previously in \cite{bariev1991integrable, fendley2003lattice,  zadnik2020folded, zadnik2020folded2, Xavier103.085101, pozsgay2021integrable}.
We uncover emergent integrability of the slow hole dynamics in that regime and demonstrate its equivalence to the constrained XXZ model of Alcaraz and Bariev \cite{Alcaraz1999}. Moreover, following \cite{PhysRevLett.127.167203}, we investigate the salient consequences of additional short-range fermion interactions in this lattice gauge theory. While for the repulsive case these interactions can stabilize the Mott state at the hole filling $\nu^h=1/3$, in the case of attraction we predict the phenomenon of clustering of holes into large conglomerates, whose hopping scales exponentially with their length.

Our work illustrates how spontaneous breaking of translation symmetry gives rise to deconfinement of excitations in one-dimensional systems and highlights the salient properties of deconfined fractionalized domain walls including integrability that emerges at low energies.  Our predictions can be probed in quantum simulators of $\mathbb{Z}_2$ gauge theories coupled to dynamical gapless matter, whose design has been recently initiated in \cite{Barbiero2018, Schweizer2019, Gorg2019, ge2020approximating, Wang_2022}.

The paper is organized as follows: In Sec. \ref{sec:dec}, we briefly discuss various features of our model and show how holes act as domain walls between vacua that spontaneously break translational symmetry, which we argue is responsible for deconfinement of hole excitations. We then deduce the effective Hamiltonian in the limit of strong string tension, which we show to be integrable, in Sec. \ref{sec:strong}. We perform time evolution of single and two-hole states, to support our theoretical argument, in Sec. \ref{sec:dynamics}. We then enrich the model in Sec. \ref{sec:short} by adding a short-range interaction term to the Hamiltonian and investigate its main physical consequences. Finally, we present our conclusion and highlight scope for future study in Sec. \ref{sec:conc}.

\section{Deconfined holes and their interactions}
\label{sec:dec}
\begin{figure}
    \centering
    \includegraphics[width = \columnwidth]{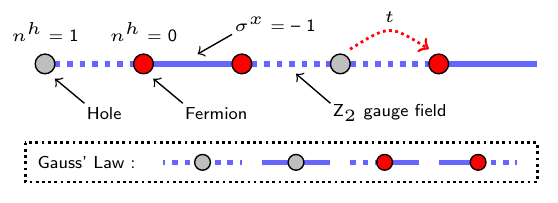}
    \caption{Holes (fermions) in grey (red) occupy sites and interact with the $\mathbb{Z}_2$ gauge fields defined on links. All physical configurations must satisfy the Gauss law.
    }
    \label{fig:model_zigzag}
\end{figure}

Our starting point is a one-dimensional chain with single-component fermions $c_i$ living on sites and $\mathbb{Z}_2$ Ising gauge fields defined on links, see Fig \ref{fig:model_zigzag}. The Hamiltonian governing the quantum system is 
\begin{equation}
    \label{1dH}
    \mathcal{H} = - t\sum_{ i } (c^{\dagger}_{i} \sigma^z _{i + 1/2} c_{i+1}  + \text{h.c.}) - h \sum_i \sigma^x_{i+1/2} - \mu \sum_{ i} n^f_i ,
\end{equation}
where $n^f_i = c^{\dagger}_ic_i$. The first term couples the fermions to the $\mathbb{Z}_2$ gauge fields via the Ising version of the Peierls substitution while the second term induces transitions for the gauge Ising spins. Finally, the chemical potential term tunes the ground state density of fermions whose total number is conserved.

The model \eqref{1dH} exhibits local $\mathbb{Z}_2$ gauge invariance with generators $G_i = \sigma^x_{i-1/2} (-1)^{n^f_i} \sigma^x_{i+1/2}$. Choosing eigenvalues $G_i = \pm 1$ gives rise to independent sectors of the Hilbert space. We shall work in the "even" sector with $G_i = +1$ for all sites. This choice corresponds to absence of static charges, i.e., all $\mathbb{Z}_2$ charges are carried by dynamical fermion matter.

On an infinite chain, we now introduce non-local hole creation and annihilation operators
\begin{equation}
\label{holedef}
h^\dagger_i = c_i \prod_{j \geq i} \sigma^z_{j+1/2}, \quad
h_i = c^\dagger_i \prod_{j \geq i} \sigma^z_{j+1/2}.
\end{equation}
Here the semi-infinite gauge string ensures that these operators are $\mathbb{Z}_2$ gauge-invariant.
The holes have fermionic statistics, satisfying the usual anti-commutation relations, as can be seen from their definition. 
By introducing hole number operators $n^h_i = h^{\dagger}_i h_i = 1 - n^f_i$ we can rewrite the $\mathbb{Z}_2$ gauge generators in terms of holes as $G_i = \sigma^x_{i-1/2} (-1)^{1 - n^h_i} \sigma^x_{i+1/2}$. 
On a closed chain of length $L$, the Gauss law condition $G_i=1$ then ensures $\prod_i (-1)^{ n_i^h}=(-1)^L$. As a result, on closed chains of an even and odd length the number of holes must be even and odd, respectively. This also implies that in a closed geometry one can add and destroy holes only in pairs, but not individually \footnote{On a closed chain the definition of the hole operators \eqref{holedef} is not gauge-invariant, but we can introduce a gauge-invariant creation operator of a pair of holes $h_i^\dagger h_j^\dagger=c_ic_j \prod_{i\leq k \leq j} \sigma^z_{k+1/2}$.}.
On the other hand, on a finite open chain which ends with links, the Gauss law does not constrain the total parity of holes and individual holes can be created and removed by applying the operators \eqref{holedef}. 

\begin{figure}
    \centering
    \includegraphics[width = \columnwidth ]{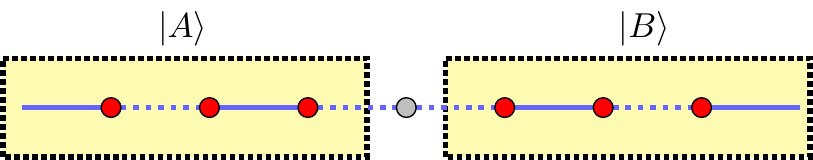}
    \caption{$\ket{A}, \ket{B}$ represent the two hole vacua with spontaneously broken translation symmetry. Notice how a single hole acts as a domain wall between the two hole vacua.}
    \label{fig:ssb}
\end{figure}

On an infinite chain, our model has two degenerate hole vacuum states which are annihilated by all $h_i$ operators. Both of these states are completely filled with fermions, but differ in the location of the electric strings that occur at odd and even links, respectively \footnote{The same is true for a finite chain with even number of sites.}. The two ground states spontaneously break translation symmetry of the Hamiltonian and holes form domain walls between the two vacua, see Fig. \ref{fig:ssb}

Since the two ground states are degenerate in energy, domain walls should be deconfined. To see that explicitly,
it is possible to express the Hamiltonian \eqref{1dH} completely in terms of the hole operators. On an infinite chain one finds the Hamiltonian to be \cite{PhysRevLett.127.167203} 
\begin{equation}
    \label{hamiltonian_holes}
    \mathcal{H} = - t \sum_i (h_i h_{i+1}^{\dagger} +  \text{h.c.})  - h \sum_i (-1)^{\sum_{j > i} 1-n^h_j}. 
\end{equation}

The details of the derivation can be found in Appendix \ref{AppA}, where we also discuss how the Hamiltonian changes on a closed chain. 
The last term in the Hamiltonian mediates an infinite-range potential between two holes. The potential has a zig-zag form which alternates between the values -$2h$ and $0$ for the odd and even distances, respectively. As a result, the two holes are deconfined and free to spread far away from each other in the absence of other holes. This is in stark contrast to the original fermionic particles which are confined due to an attractive potential that scales linearly with distance. The deconfined nature of holes is a consequence of spontaneous symmetry breaking of a global symmetry, which is a generic mechanism for fractionalization in one-dimensional systems. 

One can eliminate the $\mathbb{Z}_2$ gauge redundancy and rewrite the model \eqref{1dH} in terms of gauge-invariant spin $1/2$ degrees of freedom residing on links of the lattice \cite{borla2020confined}. In this formulation the Hamiltonian takes the local form
\begin{equation}
\label{spin_ham}
    \mathcal{H}=-\frac{t}{2} \sum_{i}\left(1-X_{i-1/2} X_{i+3/2}\right) Z_{i+1/2}-h \sum_{i} X_{i+1/2} ,
\end{equation}
where $X$ and $Z$ are $\mathbb{Z}_2$ gauge-invariant Pauli operators. In this formulation the original fermion particles are interpreted as domain walls between $X$-polarized regions. The holes thus correspond to absence of domain walls and appear on sites surrounded by links that are in the same eigenstate of the $X$ operator. The hole creation operator \eqref{holedef} can be expressed in terms of the gauge-invariant spin operators as
    $h^\dagger_i = -(X_{i-1/2}+X_{i+1/2})/2 \prod_{j \geq i} Z_{j+1/2}$ 
The details of the derivation can be found in Appendix \ref{AppB}.
The non-local character of this mapping is a mathematical manifestation of the fractionalized nature of holes.  Only pairs of holes can be created by local gauge-invariant operators.


\section{The limit of strong string tension} \label{sec:strong} In the limit where the string tension $h$ is much larger than all other energy scales in the problem, dimers of fermions (connected with the unit length electric strings) emerge as the relevant low-energy degrees of freedom \cite{borla2020confined}. In the hole picture, this corresponds to the sector where consecutive holes are always separated by odd distances.

At second order of perturbation theory in the hopping parameter $t$, the dynamics of holes in this sector is governed by the effective Hamiltonian becomes, as shown in Appendix \ref{AppC}
\begin{equation}
    \label{eff_H_2}
    \mathcal{H}_{\text{eff}} = -t_{\text{eff}} \sum_i (h_{i-1}^{\dagger}(1-n^h_i) h_{i+1} + \text{h.c.}) + U_{\text{eff}} \sum_i n^h_i n^h_{i+1}
\end{equation}
with $U_{\text{eff}} = t^2/h = 2 t_{\text{eff}}$. We observe that in this regime holes always hop by two sites. The factor $(1-n^h)$ inhibits hopping between next-nearest sites if the intermediate site is already occupied with a hole. This type of correlated hopping was first investigated by Bariev \cite{bariev1991integrable} who demonstrated its integrable nature, for related recent studies see \cite{zadnik2020folded, zadnik2020folded2, pozsgay2021integrable}. In addition to the Bariev's hopping, a nearest neighbour repulsion $U_{\text{eff}}$ is induced between the holes by the second-order perturbation theory.

\begin{figure}[t]
    \centering
    \includegraphics[width = \columnwidth]{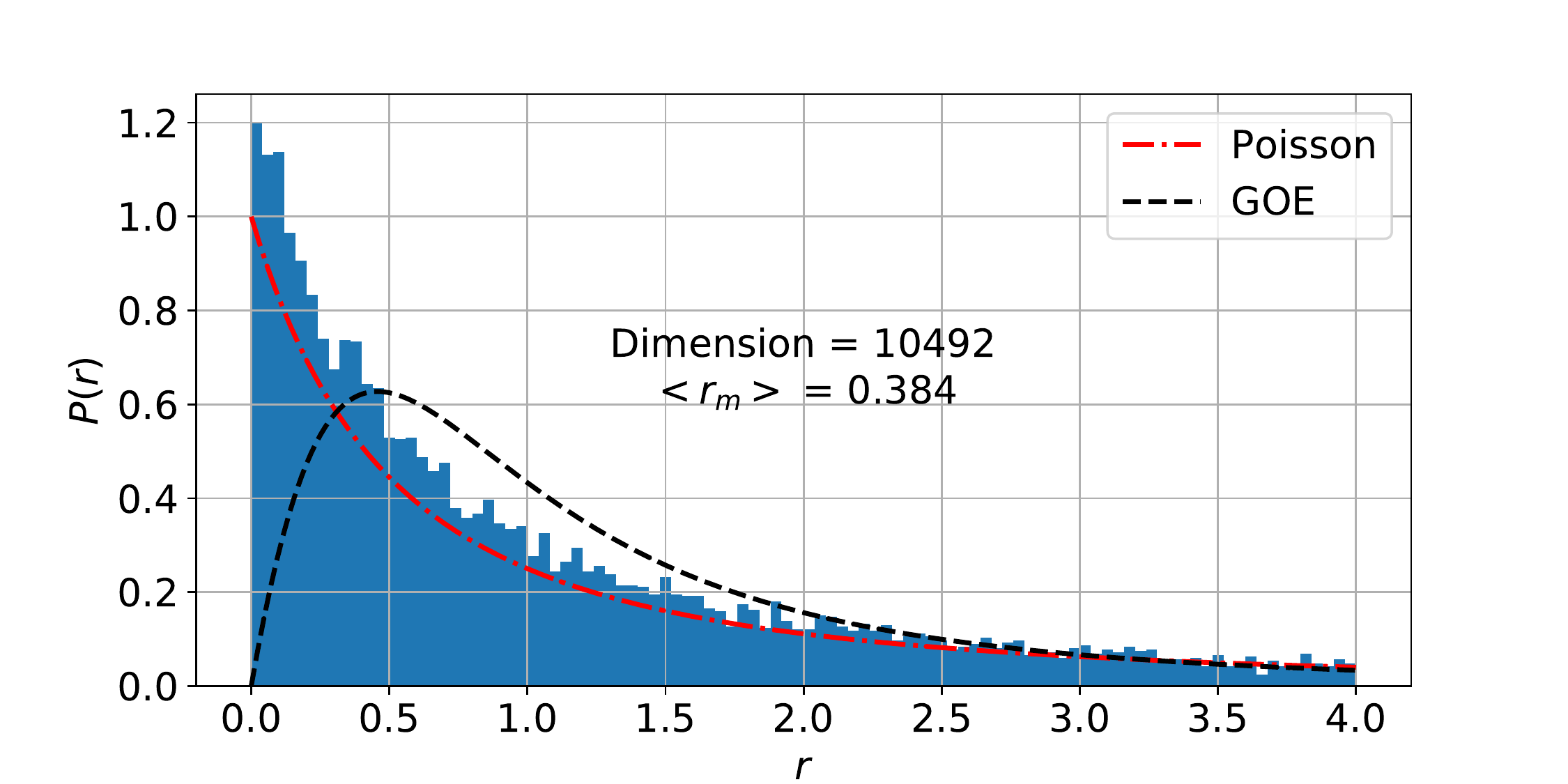}
    \caption{Probability distribution of ratios $r$ of consecutive energy differences for the Hamiltonian \eqref{eff_H_2} matches well with the integrable distribution. The average $\langle r_m\rangle =\langle \min(r,1/r) \rangle=0.384$ is close to the integrable value 0.386 \cite{Oganesyan2007}. After resolving all symmetries, ED is implemented on a closed chain of 31 sites with 15 holes.}
    \label{fig:integrability_mapping}
\end{figure}

As argued above, in the strongly-coupled regime holes always hop between next-nearest neighbour sites. Thus on an open chain holes hop independently on even- and odd-numbered sublattices. This at first sight suggests that we have two $U(1)$ conservation laws instead of just one, namely $N_e = \sum_{i \in \text{even}} n^h_i$ and $N_o = \sum_{i \in \text{odd}} n^h_i$ are separately conserved \footnote{For a closed chain the above conclusion remains valid as long as it has an even number of sites - in case of an odd number of sites, separate sublattice number conservations do not hold.}. Note however that $N_e$ and $N_o$ are not independent. Since in the investigated sector consecutive holes are always odd distance apart, the occupation of two sublattices must be essentially the same. As a result, the two $U(1)$ global symmetries are intertwined and not independent.

Our numerical exact diagonalization (ED) investigation \cite{Quspin1,Quspin2} of energy level statistics \cite{Oganesyan2007, LevelStats}, presented in Fig. \ref{fig:integrability_mapping}, reveals integrability of the effective Hamiltonian \eqref{eff_H_2}. We now demonstrate that the Hamiltonian \eqref{eff_H_2} is equivalent to the integrable constrained XXZ model introduced in \cite{Alcaraz1999}.

To demonstrate the mapping, we first define the dimer creation and annihilation operators that act on links of the lattice:
$
b^\dagger_{i+1/2}=h_{i} h_{i+1}
$
and
$
b_{i+1/2}=- h^\dagger_{i} h^\dagger_{i+1}
$.
The dimers do not behave strictly like point-like bosons because on neighbouring links they satisfy the following commutation relation
\begin{equation} \label{com}
[b_{i-1/2}, b^\dagger_{i+1/2}]= - h^\dagger_{i-1} h_{i+1}.
\end{equation}
This commutator indicates that the Hilbert space where dimer operators act has constraints. Indeed, since the dimers are made of single-component fermions, no nearest-neighbour links can be simultaneously occupied with dimers.
Consider now the correlated hopping term of holes in Eq. \eqref{eff_H_2}. Whenever a hole hops by two sites, a dimer hops in the opposite direction between neighbouring links. By using the definitions above and the commutation relation \eqref{com}, it is straightforward to show that the hopping term can be rewritten in terms of dimer operators as
$H_{\text{hop}}=-t_{\text{eff}} \sum_{i} P_1 (b_{i+1/2}^{\dagger}  b_{i-1/2}+\text { h.c. })P_1$,
where $P_1$ denotes a projector that inhibits (i) multiple dimer occupation of any link of the lattice and (ii) simultaneous occupation of dimers on neighbouring links. 

\begin{figure}[t]
    \centering
    \includegraphics[width=0.5\textwidth]{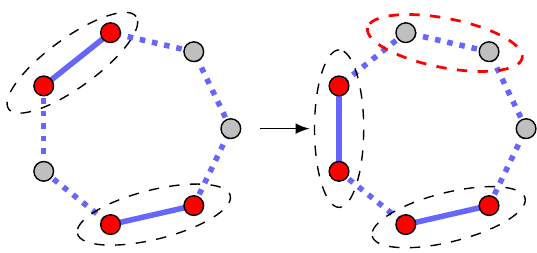}
    \caption{Towards the mapping between the hole and dimer interactions: on a closed chain, whenever a new nearest neighbour pair of holes emerges (marked in red on right), a new corresponding pair of next-nearest neighbour dimers appears. The dimers are highlighted in black.}
    \label{fig:map_int}
\end{figure}

Now we turn to the nearest neighbour interaction term between holes in the Hamiltonian \eqref{eff_H_2}. Can we rewrite it in terms of the dimer operators? At first sight, it appears to be impossible because the interaction energy density that is proportional to $n_i^h n_{i+1}^h$ cannot be expressed in terms of the dimer degrees of freedom only. Note however that on a closed chain the number of the nearest-neighbour holes is complementary to the number of the next-to-nearest dimers, see Fig. \ref{fig:map_int}. Given that, the nearest neighbour repulsion between the holes can be rewritten as the next-nearest neighbour repulsion between the dimers.

All together, (up to an unimportant energy shift) the correlated hopping model \eqref{eff_H_2} is equivalent
to the constrained model of bosonic dimers
$H=-t_{\text{eff}} \sum_{i} P_1 \Big[ (b_{i+1/2}^{\dagger}  b_{i-1/2}+\text { h.c. })- 2 n^B_{i-1/2} n^B_{i+3/2} \Big] P_1$.
After employing the standard relation between spin $1/2$ operators and hard-core bosons, we recognize  the constrained XXZ model of \cite{Alcaraz1999}. As further evidence of this equivalence, we found that the energy spectra of the \eqref{eff_H_2} and the constrained XXZ chain, $m_i \equiv E_{i}-E_{0}$, where $E_0$ is the lowest energy eigenvalue, agreed very well numerically.

\begin{figure*}[t!]
    \centering
    \includegraphics[width = 2\columnwidth]{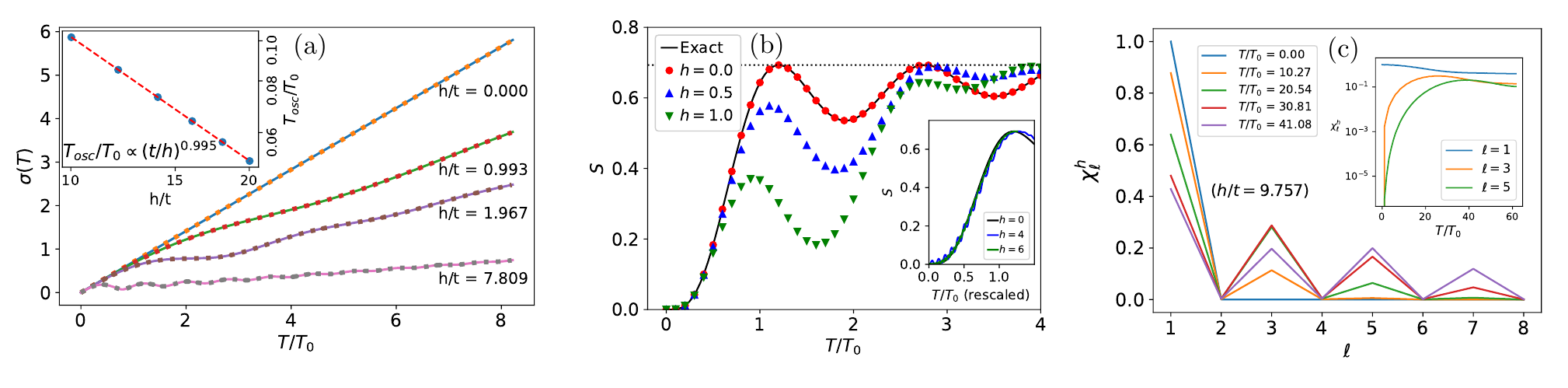}
    \caption{(a) Dynamics of holes : The solid lines denote the standard deviation obtained from the ED while the dotted lines were  computed by solving numerically Eq. \eqref{eq:time_dep_se}. For $h = 0$, we obtain an excellent agreement with \eqref{var_exp}. As $h$ is increased, we observe an oscillatory behavior on top of the overall linear growth. In the inset, the time period of the oscillations is plotted (blue dots), which decays as $h^{-1}$ for large $h$. (b) The hole entanglement entropy under a bipartition from TEBD: The solid black line is the analytical result at $h=0$. The dotted line represents $S = \ln 2$. The inset shows that under rescaling time $T$ by $t/(2h)$ the curves for $h\gg t$ collapse onto the one for $h = 0$ as expected from arguments in the main text. (c) The density-density correlator $\chi^h_l$ of a state with two holes localized nearby at $T=0$.When $h \gg t$, the holes prefer to remain odd distance apart. In the inset we plot $\chi^h_\ell(T)$ for fixed $\ell$. The likelihood of $\ell = 1$ progressively decreases while that of other odd $\ell$'s correspondingly increase, until they saturate. ED performed on chain with 19 sites.  }
\label{fig:time_evol}
\end{figure*}

The strong string tension effective theory can be also investigated in sectors containing dimers of non-minimal length. In such sectors holes are not necessarily odd distance apart and the effective Hamiltonian \eqref{eff_H_2} is not valid. In the spin formulation \eqref{spin_ham}, the effective Hamiltonian applicable in all sectors  has been computed in \cite{PhysRevLett.124.207602}. In the rotated basis $X\leftrightarrow Z$ it reads 
    $\mathcal{H}_{\text{eff}} = -t_{\text{eff}} \sum_i [Z_{i-1/2} \mathcal{P}_{i-1/2,i+5/2} ( S^+_{i+1/2} S^-_{i+3/2} + \text{h.c.} )  - Z_{i-1/2}Z_{i+1/2}Z_{i+3/2}/2]$,
where $S^{\pm}_{i+1/2}=(X_{i+1/2}\pm iY_{i+1/2})/2$ and $\mathcal{P}_{\alpha \beta}=(1+Z_\alpha Z_\beta)/2$ projects out states with opposite Z-eigenvalues on links $\alpha$ and $\beta$.
In the shortest dimer sector, this Hamiltonian reduces to the constrained XXZ model which as we demonstrated above is equivalent to the local correlated-hopping hole Hamiltonian \eqref{eff_H_2}. Note, however, that since the full effective spin Hamiltonian is made of products of odd number of spin operators, it appears that beyond the shortest dimer sector it is impossible to rewrite this Hamiltonian in terms of fractionalized holes in a local form.

\section{Hole dynamics} \label{sec:dynamics}

We now turn our attention to the time evolution of a quantum state in which a single hole is fully localized at site $m=0$ at time $T = 0$. A general single-hole state may be written as
    $\ket{\Psi} = \sum_m \psi_m h^\dagger_m \ket{0} \equiv \sum_m \psi_m \ket{m}$,
where $\ket{0}$ denotes a vacuum of holes, i.e. a state fully filled with fermions \footnote{As emphasized, there are two vacua that differ by the pattern of the electric strings. Note, however, that these two vacua are in fact related by a unitary transformation and so in the following it suffices to consider one of them as the vacuum state to be henceforth denoted as $\ket{0}$.}. In order to follow the time evolution of this state, one needs to solve the time-dependent Schr\"odinger equation, which (as we show in Appendix \ref{AppE}) for this case reads
\begin{equation}
    \label{eq:time_dep_se}
   i \partial_T \psi_m= -t(\psi_{m+1} + \psi_{m-1}) +  h (1+(-1)^{m+1} ) \psi_m 
\end{equation}
revealing an effective zig-zag potential. 
Consider first the case with $h = 0$, where the hole is free with the dispersion relation
 $   E(k) = - 2 t \cos k$.
As a result, the time-evolved state is simply given by
    $\psi_m(T) = \int_{-\pi}^{\pi} e^{2 i t T \cos k} e^{ikm} dk/(2 \pi)=J_m\left(\frac {T} {T_0}\right)$,
where $J_m(x)$ denotes the Bessel function of the first kind and $T_0=(2t)^{-1}$. To quantify the spreading of the hole in time, we compute the standard deviation (SD) of the hole from its original site
\begin{equation} \label{var_exp}
    \sigma (T)=\sqrt{\langle x^2 \rangle -  \langle x \rangle^2}   = \frac{T}{\sqrt{2} T_0},
\end{equation}
the hole spreads linearly in time with the rate controlled by the hopping parameter $t$.
Now we investigate how the spreading of a hole is affected by a finite string tension $h$. 
Fig. \ref{fig:time_evol} (a) reveals that the dynamics slows down as $h$ is increased. Moreover, on top of the linear growth  we observe damped oscillations whose frequency increases as $h$ grows.

Here we attempt to understand analytically the salient features in the limit $h\gg t$. First, the spectrum of the Schr\"odinger equation \eqref{eq:time_dep_se} forms two bands in the halved Brillouin zone with energies
    $E_\pm(k) = h \pm \sqrt{ (2t \cos k)^2 + h^2}$.
The wave function at site $n$ can be expressed as 
    $\Psi_n(T) =  \int_{-\pi/2}^{\pi/2} dk/(2 \pi) (c_{k}^{-} \phi_-(k,n) e^{-iE_- T} + c_{k}^+ \phi_+(k,n) e^{-iE_+ T})$,   
where $\phi_{\pm}(k)$ are the eigenfunctions and the coefficients $c_{k}^{\pm}$ are chosen to ensure that the hole is localized at $n=0$ at $T=0$. In the limit of large $h$, we have $E_- \approx -2T_s^{-1} \cos^2 k, \, E_+ \approx 2h + 2T_s^{-1} \cos^2 k$, where we introduced a slow time scale $T_s = h/t^2$. Thus the wave function becomes
    $\Psi_n(T) = f_n(T/T_s ) + e^{2ihT} g_n( T/T_s)$.
This form makes it manifest that the rapidly-oscillating factor $e^{2ihT}$ is responsible for the oscillations observed in Fig. \ref{fig:time_evol} (a). As a result, in the large-$h$ regime, the time scale of these oscillations scales as $h^{-1}$. The inset of Fig. \ref{fig:time_evol} (a) confirms this prediction.

We now study the time evolution of the entanglement entropy (EE) of the single-hole state investigated above under a bipartition cut at the site, where the hole is initially localized, refer to Appendix \ref{AppF} for details. At time $T=0$ we start from a product state, so the EE is vanishing. Since the hole is a single particle excitation, the corresponding EE is bounded by $\ln 2$ \cite{jia2008entanglement}. In Fig. 3 (b), we present numerical TEBD results together with the analytical  prediction at $h=0$, presented in Appendix \ref{AppF}. As expected, we find that the hole entanglement growth slows down as $h/t$ increases.  Under rescaling time $T$ by a factor of $t/2h$ the EE growth at $h\gg t$ collapses to the $h=0$ curve, see inset of Fig. \ref{fig:time_evol} b. This is because in the $h \gg t$ limit, the effective model \eqref{eff_H_2} describes pure hopping of a single hole with a time scale $T_s = h/t^2$, while \eqref{hamiltonian_holes} describes a similar kind of model with time scale $T_0 = (2t)^{-1}$ when $h = 0$. Thus one would expect the spread of entanglement to evolve similarly if one rescales the time by the appropriate factor.

\begin{figure*}[t!]
    \centering
    \includegraphics[width = 2\columnwidth]{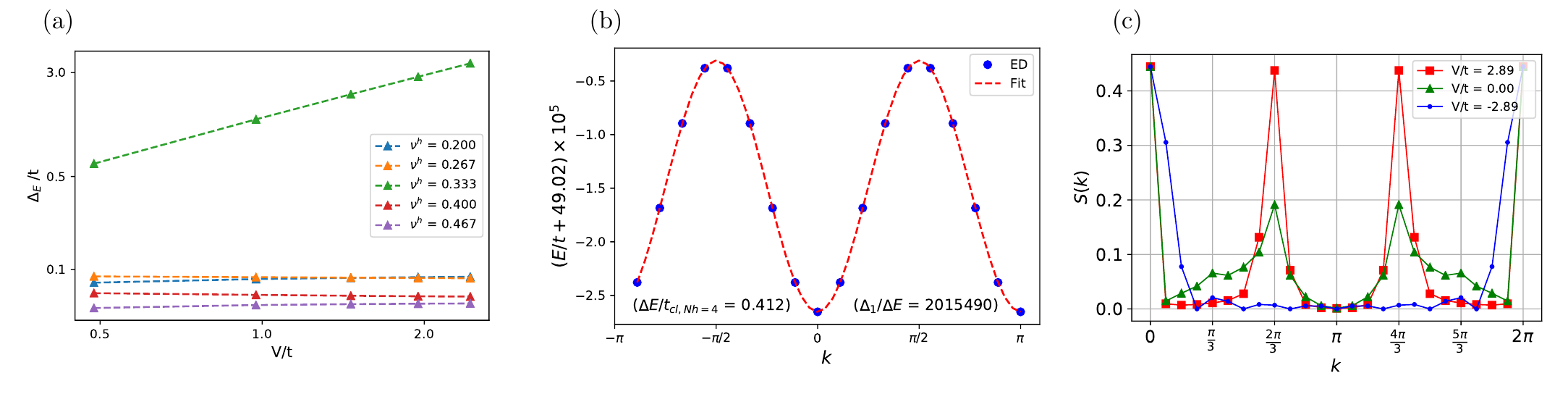}
    \caption{(a) Charge gap for the model with $t=1.027$, $h=9.09$: At the hole filling $\nu^h=1/3$ the Mott gap scales linearly with the interaction strength $V$. Away from the filling $\nu^h=1/3$, the gap is negligible. (b) Low-energy states of four holes for $t=1$, $h= 10.021$ and $V = -1.04$ : The lowest band exhibits a cosine dispersion relation $E(k) = - 2 t_{cl, N_h=4} \cos 4k$ with the width set by $t_{cl, N_h=4} \sim t_{\text{eff}}^{4}/V^{3}$.  $\Delta E$ refers to the width of the lowest band, while $\Delta_1$ refers to the energy difference between the top of the lowest band and the bottom of the next band. It is apparent that these bands are well-separated. (c) Hole structure factor at $\nu^h = 1/3$ : Sharp peaks around $k = 2\pi/3$ and $4\pi/3$ indicate the formation of a Mott solid in the repulsive regime, which disappear for $V < 0$, implying restoration of translational symmetry of the ground state.These calculations were implemented on a closed chain with 24 sites, with $t = 1.037$ and $h = 10.021$ using ED.} 
\label{fig:short_range}
\end{figure*}

We next perform the ED time evolution of a pair of holes to shed more light on their deconfined nature. We begin with a state in which the two holes (of top of a hole vacuum) are localized at neighbouring sites initially at $T = 0$. We compute the density-density correlator $\chi^h_\ell=\sum_k \langle n^h_k n^h_{k+\ell} \rangle$ which measures the likelihood of the separation between the two holes, see Fig. \ref{fig:time_evol} (c). As expected, holes spread away from each other and in the large $h$ limit prefer to be an odd distance apart.
In contrast, the corresponding computations of the fermionic density-density correlator $\chi^f_\ell=\sum_k \langle n^f_k n^f_{k+\ell} \rangle$ for a pair of fermions (on top of the fermionic vacuum) reveals that they  remain closely confined together.

Above arguments illustrate the deconfined nature of the lattice hole as a domain wall between the two vacua fully filled with fermions. At finite density of holes, however, the translation symmetry of the ground state is restored and lattice holes become confined. Indeed, at $h\ne 0$ one observes that the hole-hole correlation function decays exponentially. As a result, at any finite hole filling the lattice operator creating a hole does not coincide with the annihilation operator of the emergent deconfined fermionic excitation of the Luttinger liquid field theory discussed in \cite{borla2020confined}.


\section{Short-range interactions}
\label{sec:short}
To illustrate how new patterns of translation symmetry breaking and deconfinement of holes can emerge at lower fillings, we now go beyond the pure $\mathbb{Z}_2$ gauge interactions and enrich the model by adding a short-range density-density nearest neighbour interaction term \cite{PhysRevLett.127.167203} 
\begin{equation}
    \label{nnHam}
    \mathcal{H}_{\text{nn}} = \mathcal{H} + V \sum_i n^f_i n^f_{i+1}
\end{equation}
where $\mathcal{H}$ is defined in \eqref{1dH} . In terms of the hole operators, this just corresponds to adding the term $V \sum_{ i} (1-n^h_i) (1-n^h_{i+1})$ to the Hamiltonian \eqref{hamiltonian_holes}.  

In order to gain some qualitative understanding of the interplay between the gauge and short-range forces, we consider first the static regime, where the hopping $t$ is set to zero. In this case, a bare repulsion $V>0$ inhibits nearest neighbour occupation of holes. In the ground state, for a low filling $\nu^h<1/3$, the holes will arrange themselves an odd $l>2$ distance apart such that unit-length electric strings connect the complementary fermions. On the other hand, a bare attraction $V<0$ will favor a ground state in which the holes are clustered together into a large conglomerate. 

In order to understand the case with non-zero hopping $t$ analytically, we look here at the strong tension limit $h \gg |V|, t$. The short-range interaction term gives rise to a simple modification of the effective interaction $U_{\text{eff}}=t^2/h\to U_{\text{eff}}+V$ in the effective Hamiltonian \eqref{eff_H_2}.
Since in the strong tension limit $|V|\gg t_{\text{eff}}, U_{\text{eff}}$, effectively the short-range interaction imposes a constraint on the low-energy Hilbert space. In the case of bare repulsion ($V > 0$) and low filling ($\nu^h < 1/3$), the low-energy constraint imposed is $n^h_i n^h_{i+1} = 0$, so that the effective model reduces to 
\begin{equation}
   \label{eq:low_filling_pos_V}
    \mathcal{H}_{\text{eff}} = - t_{\text{eff}} \sum_i \mathcal{P}_2( h_{i-1}^{\dagger}  h_{i+1} + \text{h.c.} )\mathcal{P}_2,
\end{equation}
where $\mathcal{P}_2$ is a projector that excludes holes from occupying a distance of less than or equal to two.  
Precisely at $\nu^h = \frac{1}{3}$, holes occupy every third lattice site, forming a Mott state with a gap of order $V$. This gap was calculated with ED as  
 $   \Delta_E(N_h) = \frac{1}{2} ( E^{(0)}_{N_h+2 }+E^{(0)}_{N_h-2} - 2 E^{(0)}_{N_h})$
where $E^{(0)}_{N_h}$ refers to the ground state energy of chain with $N_h$ holes.  At fillings away from $\nu^h = \frac{1}{3}$ we detected no sizable energy gap above the ground state which is consistent with a Luttinger liquid behavior, Fig. \ref{fig:short_range} (a). On the other hand, in this regime a bare attraction ($V < 0$) that over-weights the induced repulsion $U_\text{eff}=t^2/h$ between the holes, makes a ground state in which all the holes are clustered in a single conglomerate energetically preferable. A simple perturbative estimate suggests that the double-site hopping of the hole cluster of size $N_h$ is suppressed exponentially and scales as $t_{cl, N_h}\sim t^{N_h}_{\text{eff}}/V^{N_h-1}$. As a result, clusters have the cosine dispersion $E(k) = - 2 t_{cl, N_h} \cos 4k$ which is indeed supported by ED, see Fig. \ref{fig:short_range} (b).

To shed some light on properties of the ground state under lattice translations, we measured the hole structure factor 
\begin{equation}
    \label{eq:struct-factor-def}
    S(k) = 1/L \sum_{j, l = 0}^{L-1} e^{ikl} \langle n^h_j n^h_{j+l} \rangle
\end{equation}
at filling $\nu^h = 1/3$ with the help of ED. Our results are summarized in Fig. \ref{fig:short_range} (c).
In the repulsive case, in addition to a peak at vanishing momentum we observe two additional sharp peaks at $ k = 2 \pi/3$ and $4 \pi/3$, consistent with translation symmetry breaking pattern of the Mott state described above. One also observes that upon decreasing the repulsion, the peaks become less sharp. For a sufficiently large attractive potential, these peaks eventually completely disappear indicating restoration of the translation symmetry of the ground state.

Given that the $\nu^h=1/3$ Mott state breaks translational symmetry spontaneously, additional isolated holes on top of such a state constitute domain walls between (three) different symmetry broken ground states. As a result, they are deconfined excitions with properties similar to the deconfined holes on top of fully filled vacua discussed above.\\

\section{Conclusion and outlook} \label{sec:conc}
In this paper we investigated deconfined dynamics of one-dimensional fermionic domain walls on top of a translationally spontaneously broken ground state. While we concentrated our discussion on the full fermionic filling here, similar physics should emerge at the hole fillings $\nu^h=1/3$ \cite{PhysRevLett.127.167203} and $\nu^h=1/2$ \cite{kebrivc2022confinement}, where translation symmetry breaking Mott ground states in the Ising gauge theory can be stabilized by additional short-range repulsive interactions.  Beyond the model studied here, salient features of our findings should be applicable to other one-dimensional (spin) chains and quasi-one-dimensional (spin) ladders with translationally broken ground states. For example, the soliton-induced deconfinement discovered in the $\mathbb{Z}_2$ gauge theory coupled to fermions on the Creutz-Ising ladder \cite{PhysRevX.10.041007} is rooted in translation symmetry breaking. Another closely related system, where this mechanism might be applicable at some special fillings, is the theory on a triangular ladder investigated in \cite{PhysRevB.105.245105}. Maybe, in some form, ideas presented here can be extended to higher dimensions, where deconfinement of excitations and associated topological order originate from spontaneous breaking of higher-form symmetries \cite{gaiotto2015generalized, PhysRevB.99.205139}.


\begin{acknowledgements}
We would like to acknowledge useful discussions on a related project with Luca Barbiero, Fabian Grusdt and Matjaz Kebri\v c. We are grateful to Fabian Grusdt and Matjaz Kebri\v c for pointing out the confined nature of lattice holes at finite filling of holes. We are grateful to Alvise Bastianello for discussions about the relation between one-dimensional deconfinement and translation symmetry breaking.
Our work is funded by the Deutsche Forschungsgemeinschaft (DFG, German Research Foundation) under Emmy Noether Programme grant no.~MO 3013/1-1 and under Germany's Excellence Strategy - EXC-2111 - 390814868. This work is supported by
Vetenskapsr\aa det (grant number 2021-03685).
    
\end{acknowledgements}

\bibliography{library}

\clearpage

\appendix


\begin{widetext}


\section{Hamiltonian in terms of hole operators} \label{AppA}
Here we rewrite the Hamiltonian \eqref{1dH} in terms of the non-local gauge-invariant hole creation and annihilation operators introduced in Eq. \eqref{holedef}.

First we consider an infinite chain.
Substituting $ c_i = h_i^{\dagger} \prod_{j \geq i} \sigma^z_{j+1/2}$  in the first term of the Hamiltonian \eqref{1dH}, we get 
\begin{equation}
 - t \sum_i \Big[ h_{i} (\prod_{j \geq i} \sigma^z_{j+1/2})  \sigma^z_{i+1/2} (\prod_{j \geq i+1} \sigma^z_{j+1/2})  h_{i+1}^{\dagger} + \text{h.c.} \Big]. 
\end{equation}
 Since all $\sigma^z$ operators square to one, the hopping term simplifies 
\begin{equation}
    \label{hop_hole}
    \mathcal{H}_t = - t \sum_i (h_i h_{i+1}^{\dagger} +  h_{i+1}h_{i}^{\dagger}).
\end{equation}

In order to rewrite the electric term in terms of the hole operators, we will  make use of the Gauss law constraint. 
In particular, we consider an infinite product of the $\mathbb{Z}_2$ generators on sites $j>i$. Given that we work in the even gauge theory, $\prod_{j > i} G_j=1$. Except for the link $i+1/2$, it is clear that there is always two factors of $\sigma^x$ operators acting on every link, which just square to one. This leaves us with the identity $\sigma^x_{i+1/2}= (-1)^{\sum_{ j > i} 1-n^h_j} $. As a result, the electric term becomes
\begin{equation}
    \label{electric_hole_inf}
    \mathcal{H}_h = -h \sum_i (-1)^{\sum_{j > i} 1-n^h_j}.
\end{equation}
Then collecting all the terms together, we end up with the Hamiltonian presented in the main text
\begin{equation}
    \label{ham_hole_inf}
    \mathcal{H} = - t \sum_i ( h_i h_{i+1}^{\dagger} +  \text{h.c.})   - h \sum_i (-1)^{\sum_{j > i} 1-n^h_j}.
\end{equation}

On a closed chain, inserting a string of $\mathbb{Z}_2$ generators on all the sites, leads to the parity constraint $P = (-1)^{\sum_j 1 - n^h_j} = 1$ and cannot be used to fix the electric field in terms of holes. To circumvent this issue, we choose an arbitrary lattice site $b$ and assign it to be the last one in the product of the $\mathbb{Z}_2$ generators. We now can express the electric field on a link as $\sigma^x_{i+1/2}= \sigma^x_{b+1/2} (-1)^{\sum_{ b \geq j > i} 1-n^h_j}$. Using this identity we end up with the Hamiltionian

\begin{equation}
\label{ham_hole_closed}
    \mathcal{H} = - t \sum_i (h_i h^{\dagger}_{i+1} +  \text{h.c.})  - h \sigma^x_{b+1/2} \sum_i (-1)^{\sum_{b \geq j > i} 1-n^h_j}.
\end{equation}
On a closed chain individual gauge-invariant hole operators $h_i$ and $h_i^\dagger$ are ill-defined. Notwithstanding, the bilinear $h_i^\dagger h_{i+1}$ and the hole occumpation number $n_i^h$ that appear in the Hamiltonian are well-defined and gauge-invariant.

\section{Hole operators in terms of gauge invariant spin$-1/2$ operators}
\label{AppB}

We start from the definitions of the gauge-invariant Pauli matrix operators introduced in \cite{borla2020confined}
\begin{equation}
    \label{xz_def}
    X_{i+1/2} = \sigma^x_{i+1/2}, \qquad Z_{i+1/2} = - i \tilde{\gamma}_i \sigma^z_{i+1/2} \gamma_{i+1},
\end{equation}
where we introduced the Majorana operators $\gamma_i = c^\dagger_i + i c_i$ and $\tilde{\gamma}_i = i(c^\dagger_i - c_i)$. Equivalently, $c_i = (\gamma_i + i \tilde{\gamma_i})/2$ and $c_i^\dagger = (\gamma_i - i \tilde{\gamma_i})/2$. In terms of the Majorana operators, $(-1)^{n^f_i} = i \tilde{\gamma_i} \gamma_i $. Then the Gauss law condition becomes $ i \tilde{\gamma_i} \gamma_i = X_{i-1/2}X_{i+1/2} $. 

Consider now the hole creation operator defined on an infinite chain as
\begin{equation}
    \label{hole_def}
    h^\dagger_i = c_i \prod_{j \geq i} \sigma^z_{j+1/2}.
\end{equation}
Using the definitions above, we rewrite this operator as 
\begin{equation}
    h^\dagger_i = \frac{(\gamma_i + i \tilde{\gamma_i})}{2} \prod_{j \geq i} (i \tilde{\gamma_j} Z_{j+1/2} \gamma_{j+1}).
\end{equation}
Rearranging the terms properly, we get
\begin{equation}
    \begin{split}
    h^\dagger_i =  & \frac{1}{2} (-1 + i\gamma_i \tilde{\gamma_i} )   (i \tilde{\gamma}_{i+1} \gamma_{i+1}) (i \tilde{\gamma}_{i+2} \gamma_{i+2}) \dots \prod_{j \geq i} Z_{j+1/2}
    \\ & = \frac{1}{2} (- 1 - X_{i-1/2}X_{i+1/2})(X_{i+1/2}X_{i+3/2})(X_{i+3/2}X_{i+5/2}) \dots \prod_{j \geq i} Z_{j+1/2}
    \\ & = -\frac{1}{2} (X_{i-1/2} + X_{i+1/2}) \prod_{j \geq i} Z_{j+1/2}. 
    \end{split}
\end{equation}
Note that in going from the first line to the second line, we used the Gauss law condition and the anti-commutation of the Majorana operators. It is straightforward to demonstrate that the hole creation operator that we found above has the correct commutation relation $[N^f, h_i^\dagger]=-h_i^\dagger$ with the fermion particle number $N^f=\sum_i (1-X_{i-1/2}X_{i+1/2})/2$. It then trivially follows that the hole-hole correlator is 

\begin{equation}
    h^\dagger_i h_j = \frac{1}{4} (X_{i-1/2} + X_{i+1/2}) \prod_{j \geq k \geq i} Z_{k+1/2} (X_{j-1/2} + X_{j+1/2})
\end{equation}

\section{Effective Hamiltonian at second order}
\label{AppC}

We set up the perturbation theory as follows. The degenerate space is defined by the Hamiltonian
\begin{equation}
    H_0 = - h \sigma^x_{b+1/2} \sum_i (-1)^{\sum_{b \geq j > i} (1-n^h_j)}
\end{equation}
while the perturbing Hamiltonian is
\begin{equation}
    V = - t \sum_i h^{\dagger}_i h_{i+1} + \text{h.c.}
\end{equation}

We use the Schrieffer-Wolff transformation \cite{schriefferWolf, Bravyi20112793} to obtain the second-order effective Hamiltonian

\begin{equation}
    \label{eff_H_def}
    \mathcal{H}^{(2)}_{\text{eff}} = \mathcal{P} ([S^{(1)},V] + \frac{1}{2!}[S^{(1)} ,[S^{(1)}, H_0 ] ]) \mathcal{P},
\end{equation}
where $\mathcal{P}$ is the projector into the degenerate manifold and $S^{(1)}$ satisfies $[S^{(1)},H_0] = -V$, from which it follows that
\begin{equation}
    \label{def_S1}
    \bra{\sigma} S^{(1)} \ket{\sigma '}  = \frac{\bra{\sigma} V \ket{\sigma '}}{\bra{\sigma} H_0 \ket{\sigma} - \bra{\sigma '} H_0\ket{\sigma '}}.
\end{equation}

\begin{figure}[h!]
    \centering
    \includegraphics[width=0.6\textwidth]{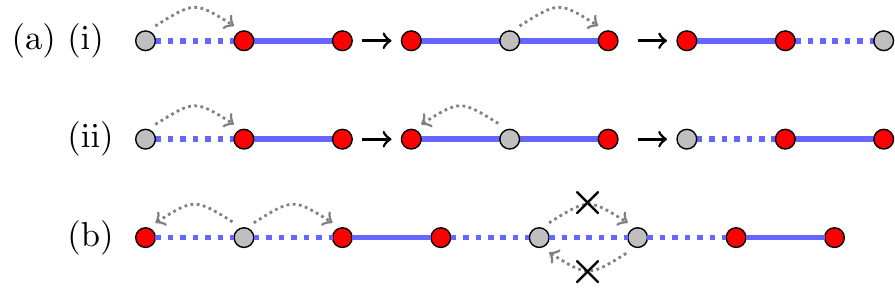}
    \caption{Second-order processes: (a)(i) demonstrates the induced hopping, while (a)(ii) displays the processes that lower the energy of the hole by $t^2/h$. (b) The crossed out lines indicate blocked virtual processes. As a result the hole on the right of the pair cannot hop to the left over the left hole (and vice versa). This blockade also implies that two out of four processes of type (a) (ii) for two consecutive holes, are forbidden. This is the origin of the second-order nearest neighbour repulsion term in the effective Hamiltonian.}
    \label{fig:deg_process}
\end{figure}

Substituting these into the above equation we get (with $\mathcal{D}$ denoting the degenerate subspace)
\begin{subequations}
\begin{align}
\begin{split}
    \mathcal{H}^{(2)}_{\text{eff}} = & \sum_{ \ket{\sigma''} \notin \mathcal{D}} \sum_{ \ket{\sigma}, \ket{\sigma'} \in \mathcal{D}} \ket{\sigma}   \frac{\bra{\sigma} V \ket{\sigma''} \bra{\sigma''} V \ket{\sigma'}}{\bra{\sigma}H_0\ket{\sigma} - \bra{\sigma''}H_0\ket{\sigma''} }   \bra{\sigma'}
\end{split}\\
\begin{split}
    = & -\frac{t^2}{2h} \sum_i ( h^{\dagger}_{i-1} (1-n^h_i) h_{i+1} + \text{h.c.}) + \frac{t^2}{h} \sum_i n^h_i n^h_{i+1}.
\end{split}
\end{align}
\end{subequations}
The second-order processes contributing to the effective Hamiltonian are shown in Fig. \ref{fig:deg_process}.

\section{Single hole dynamics}
\label{AppE}
Consider a general single hole state
\begin{equation}
    \label{state_def}
    \ket{\Psi} = \sum_m \psi_m h^\dagger_m \ket{0} \equiv \sum_m \psi_m \ket{m},
\end{equation}
where $\ket{0}$ denotes the hole vacuum. First it is straightforward to act with the kinetic term $\mathcal{H}_t = -t\sum_j (h^\dagger_jh_{j+1} + \text{h.c.})$ on this state
\begin{equation}
    \label{hopping}
    \mathcal{H}_t \ket{\Psi} = -t\sum_m \psi_m(\ket{m-1} + \ket{m+1}) = -t \sum_m (\psi_{m+1} + \psi_{m-1}) \ket{m}.
\end{equation}

Now we turn to the electric term of the Hamiltonian $\mathcal{H}_h = -h \sum_j (-1)^{\sum_{i>j}1-n^h_i}$. Firstly using the identity $(-1)^P = (1 - 2P)$ where $P$ is any idempotent operator (i.e. $P^2 = P$), we have the following equality
\begin{equation}
    \label{electric_term_eq}
    (-1)^{\sum_{i>j} (1-n^h_i)} = \prod_{i>j} (1- 2(1-n^h_i) ) = \prod_{i > j} (2n^h_i - 1),
\end{equation}
where the product follows from the fact that the number operators at different sites commute with each other. 
We can now check the action of $\mathcal{H}_h \ket{\Psi}$ of the single hole state \eqref{state_def}. For this firstly note that $n^h_i n^h_j \ket{m} = 0$ if $i \neq j$. Thus in the single hole sector we can drop all non-linear terms, i.e., 
\begin{equation}
 \prod_{i > j} (2n^h_i - 1) \to (-1)^{L-j} + 2  \underbrace{ (-1)^{L-j-1} \sum_{i > j} n^h_i}_{\equiv M^h_j}.
\end{equation}
Without loss of generality, from hereon we take $L$ to be even. The first term is independent of the position $m$ of the hole and alternates its sign as one changes the index $j$. Thus on an even-length chain, the contribution of this term can be ignored. Moreover, note now that
\begin{equation}
    M^h_j \ket{m} = \begin{cases} 
      0 & j \geq m, \\
      (-1)^{L-j-1} \ket{m} & j < m. \\
   \end{cases}
\end{equation}
Thus we end up with the following equality
\begin{equation}
    \label{electric_term_1}
    \sum_{j}M^h_j \ket{m} =  \sum_{j<m} (-1)^{L-j-1} \ket{m} = - \frac{1+(-1)^{m-m_0+1}}{2} \ket{m}
\end{equation}
where $m_0$ is the label of the first site. 

Hence the complete action of the electric term is
\begin{equation}
    \label{electric_term}
    \mathcal{H}_h \ket{\Psi} =  h (1+(-1)^{m-m_0+1} ) \psi_m \ket{m}.
\end{equation}
Putting now everything together,  we find that the time-dependent single-particle Schr\"odinger equation that governs the dynamics of the single hole sector is
\begin{equation}
    \label{time_dep_schrodinger_single_particle}
   i \partial_T \psi_m = -t(\psi_{m+1} + \psi_{m-1}) + h (1+(-1)^{m-m_0+1} ) \psi_m.
\end{equation}
\section{Single hole entanglement entropy}
\label{AppF}

\begin{figure}
    \centering
    \includegraphics[width = 0.8\textwidth]{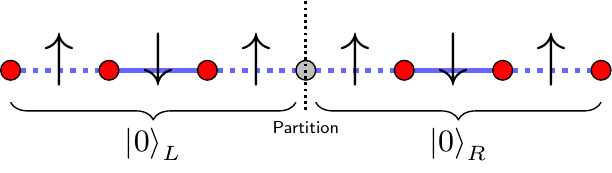}
    \caption{Bipartition of the chain: We make a cut at the site $i = 0$ where the hole is initially fully localized. In the gauge-invariant spin formulation, the $\ket{0}_{L(R)}$ refers to the left (right) anti-ferromagnetic semi-infinite domain. Up and down arrows denote the two eigenvectors of the gauge-invariant $X$ operators that act on the links of the lattice.}
    \label{fig:spin_lang_ee}
\end{figure}
We work in the gauge-invariant formulation developed in Ref. \cite{borla2020confined}, where the physical spin $1/2$ degrees of freedom live on links of the lattice. We consider an infinite chain which is partitioned in the "middle", at site labelled $i=0$. We will denote the left and right parts by "L" and "R", respectively. At $T = 0$, the hole is initially completely localized at the site $i = 0$. The corresponding initial quantum product state, denoted by $|0\rangle=|0\rangle_L |0\rangle_R$, is formed by two anti-ferromagnetic semi-infinite domains, see Fig. \ref{fig:spin_lang_ee}. We assume that $|0\rangle$ is normalized, i.e., ${}_L\braket{0|0}_L = {}_R\braket{0|0}_R = 1$. 

The time-evolved state can be written as
\begin{equation}
    \ket{\Psi(T)} = \psi_0 (T) \ket{0}_L \ket{0}_R + \ket{1}_L \ket{0}_R + \ket{0}_L \ket{1}_R,
\end{equation}
where $\ket{1}_{L(R)}$ denotes the quantum state where the hole is entirely localized on the left(right) side, respectively. In terms of $\ket{m}=h^\dagger_m |0\rangle$ defined in the previous subsection, we have
\begin{subequations}
\begin{equation}
    \ket{1}_L = \sum_{i < 0} \psi_i(T) \ket{i},
\end{equation}
\begin{equation}
    \ket{1}_R = \sum_{i > 0} \psi_i(T) \ket{i}.
\end{equation}
\end{subequations}
Now by construction, ${}_L\braket{1|0}_L = {}_R\braket{1|0}_R = 0$. Introducing,
\begin{equation}
{}_L\braket{1|1}_L = \sum_{i<0} |\psi_i|^2 \equiv p_L, \qquad {}_R\bra{1}\ket{1}_R = \sum_{i>0} |\psi_i|^2 \equiv p_R, \end{equation}
we define the following normalized states
\begin{equation}
    \ket{I}_L \equiv \frac{1}{\sqrt{p_L}} \ket{1}_L, \qquad \ket{I}_R \equiv \frac{1}{\sqrt{p_R}} \ket{1}_R.
\end{equation}

We can now compute the reduced density matrix 

\begin{equation}
\begin{split}
    \rho_L & =  tr_R \ket{\Psi(T)} \bra{\Psi(T)} \\
    & =  (p_R + |\psi_0|^2)\ket{0}_L\bra{0}_L + p_L  \ket{I}_L\bra{I}_L + \sqrt{p_L}\big( \psi_0 \ket{0}_L\bra{I}_L + \psi_0^* \ket{I}_L\bra{0}_L\big).
\end{split}
\end{equation}
Using $p_R = p_L \equiv p$ (from symmetry), $p_0 = |\psi_0|^2$ and $ 2p + p_0 = 1$ , we can write down the reduced density matrix in the matrix form 
\begin{equation}
\label{red_dm}
    \rho_L = \begin{pmatrix}
\frac{1+p_0}{2} & \psi_0^*\sqrt{p} \\
\psi_0\sqrt{p} & \frac{1-p_0}{2}
\end{pmatrix}.
\end{equation}

We can then easily see that 
\begin{equation} \label{EEan}
    S(T) = - \lambda_- \log \lambda_- - \lambda_+ \log \lambda_+,
\end{equation}
where $\lambda_\pm = (1 \pm \sqrt{p_0(2-p_0)})/2$ are the eigenvalues of the density matrix \eqref{red_dm}. For $h = 0$, the hole wave function $\psi_i(T)$ was computed in the main text. The resulting probability at site $i=0$ is $p_0 = J_0^2(T/T_0)$. Substituting this into the entanglement entropy \eqref{EEan}, one gets an analytic expression, which is compared with the data obtained from the TEBD evolution, as shown in Fig. \ref{fig:time_evol} in the main text. Note that as $T\ll T_0$, one finds $S(T) \sim (T/T_0)^4 \log (T/T_0)$ at $h=0$.

\end{widetext}

\end{document}